\documentclass{article}

\usepackage{arxiv}
\pdfoutput=1

\usepackage[utf8]{inputenc} 
\usepackage[T1]{fontenc}    
\usepackage{hyperref}       
\usepackage{url}            
\usepackage{booktabs}       
\usepackage{amsfonts}       
\usepackage{nicefrac}       
\usepackage{microtype}      
\usepackage{doi}

\usepackage{float}
\usepackage[table]{xcolor}

\usepackage{enumerate}

\usepackage{natbib} 

\usepackage{amsmath}
\usepackage{amssymb}
\usepackage{bm}
\usepackage{amsthm}

\usepackage{graphicx} 
\usepackage[font=footnotesize,labelfont=bf]{caption}
\usepackage{subcaption}

\usepackage{array}
\usepackage{booktabs}

\usepackage{cleveref}

\title{An information metric for comparing and assessing informative interim decisions in sequential clinical trials}
\date{}

\author{ \href{https://orcid.org/0000-0001-8765-0003}{\includegraphics[scale=0.06]{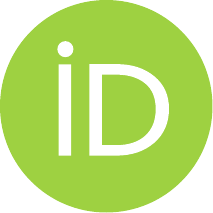}\hspace{1mm}Gianmarco~Caruso}\thanks{corresponding author: \texttt{gianmarco.caruso@mrc-bsu.cam.ac.uk}} \\
	MRC Biostatistics Unit\\
	University of Cambridge\\
	Cambridge, UK  \\
	\And
        \href{https://orcid.org/0000-0002-6513-8006}{\includegraphics[scale=0.06]{orcid.pdf}\hspace{1mm}Wlliam~F.~Rosenberger} \\
    	Department of Statistics\\
    	George Mason University\\
    	Fairfax, VA, USA \\
        \And
        \href{https://orcid.org/0000-0001-6810-0284}{\includegraphics[scale=0.06]{orcid.pdf}\hspace{1mm}Pavel~Mozgunov} \\
	MRC Biostatistics Unit\\
	University of Cambridge\\
	Cambridge, UK  \\
        \And
	\href{https://orcid.org/0000-0003-1299-9299}{\includegraphics[scale=0.06]{orcid.pdf}\hspace{1mm}Nancy~Flournoy} \\
	Department of Statistics\\
	University of Missouri\\
	Columbia, MO, USA \\
}

\renewcommand{\headeright}{Pre-print version}
\renewcommand{\undertitle}{Pre-print version}
\renewcommand{\shorttitle}{}

\hypersetup{
pdftitle={An information metric for comparing and assessing informative interim decisions in sequential clinical trials},
pdfauthor={Gianmarco~Caruso, William~Rosenberger, Pavel~Mozgunov, Nancy~Flournoy}
}

\begin{document}
\maketitle

\def\spacingset#1{\renewcommand{\baselinestretch}%
{#1}\small\normalsize} \spacingset{1}
\spacingset{1.2} 

\begin{abstract}
Group sequential designs enable interim analyses and potential early stopping for efficacy or futility. While these adaptations improve trial efficiency and ethical considerations, they also introduce bias into the adapted analyses. We demonstrate how failing to account for informative interim decisions in the analysis can substantially affect posterior estimates of the treatment effect, often resulting in overly optimistic credible intervals aligned with the stopping decision. Drawing on information theory, we use the Kullback-Leibler divergence to quantify this distortion and highlight its use for post-hoc evaluation of informative interim decisions, with a focus on end-of-study inference. Unlike pointwise comparisons, this measure provides an integrated summary of this distortion on the whole parameter space. By comparing alternative decision boundaries and prior specifications, we illustrate how this measure can improve the understanding of trial results and inform the planning of future adaptive studies. We also introduce an expected version of this metric to support clinicians in choosing decision boundaries. This guidance complements traditional strategies based on type-I error rate control by offering insights into the distortion introduced to the treatment effect at each interim phase. The use of this pre-experimental measure is finally illustrated in a group sequential trial for evaluating a treatment for central nervous system disorders.
\end{abstract}


\section{Introduction}
\label{sec:intro}

Group sequential designs (GSDs) extend traditional fixed-sample designs (FSDs) by incorporating one or more interim analyses, allowing for periodic evaluation of accumulating data in clinical trials \citep{jennison1999group}. This approach enables early termination of the trial for efficacy, when strong evidence supports the new treatment, or for futility, reducing patient exposure to ineffective interventions. By adapting to emerging results, group sequential designs can reduce the size and duration of the trial, potentially accelerating the approval of effective therapies. In recent years, GSDs have gained popularity, driven by a call for more efficient and timely evaluation of interventions. Notably, the COVID-19 pandemic underscored the need for adaptive trial methodologies capable of delivering rapid results without compromising scientific integrity \citep{stallard2020efficient,kunz2020clinical,kunz2024bayesian}.

The increased flexibility offered by GSDs comes at the cost of greater complexities in preserving the desired type-I error probability, defined in this context as the probability of incorrectly rejecting the null hypothesis at any interim look.
In fact, using the same critical value at each interim look as in a FSD leads to an inflation of the overall type-I error rate because of the repeated testing \citep{armitage1969repeated}. Various types of critical values, also referred to as stopping boundaries, have been proposed in the literature to ensure type-I error rate control in sequential settings. Notably, \cite{pocock1977group} introduced an approach that uses constant decision boundaries across all interim analyses, resulting in a relatively high probability of rejecting at an early interim analysis. 
\cite{o1979multiple} proposed boundaries that are more conservative in early analyses and become less stringent in later stages, making it more difficult to reject the null hypothesis at early analyses. Although the type-I error rate is traditionally associated with the frequentist framework, methods to control it are often employed in association with Bayesian approaches \citep{gsponer2014practical}.

While much of the methodological development in adaptive designs has focused on defining boundaries to control type-I error rates and to ensure adequate power, less attention has been given to addressing the impact that interim looks have on the estimation of treatment effects and the associated uncertainty. For example, trials with promising early results are more likely to proceed, while those with weaker effects may be stopped early for futility. This creates a bias in the estimation of the treatment effect, as early stopping tends to occur when interim results show an overestimated treatment benefit \citep{freedman1989comparison}. As a consequence, final estimates of the treatment effect can be systematically inflated, and confidence or credible intervals may not reflect the true uncertainty or may not maintain the correct coverage. \cite{flournoy2020bias} show that the bias of cumulative sample means is proportional to minus the covariance between the sum of observed outcomes and the stopping decision, which determines the sample size.
For a methodological review of the approaches proposed to mitigate this issue in point and interval estimation, the reader is referred to \cite{jennison1990statistical,jennison1993sequential} and \cite{robertson2023point,robertson2024confidence}.

Several studies have investigated the impact of informative interim decisions on the estimation of the treatment effect and the associated loss of information. In particular, \cite{marschner2021general} introduces a framework that decomposes the full likelihood into the design likelihood and the conditional-on-decisions likelihood, providing a meta-analytic perspective to assess estimation bias. \cite{tarima2024cost} examine how informative interim adaptations influence the distribution of sample-based statistics, defining the ``cost of adaptation" as the proportion of Fisher information consumed by these adaptations. 

Posterior distributions, which encapsulate all available knowledge and uncertainty about the parameter of interest after observing the data, serve as the core of all Bayesian posterior estimates and probability statements \citep{gelman1995bayesian}.
In this respect, \cite{flournoy2023posterior} advocate constructing Bayesian posteriors that incorporate the informative interim decisions made during a group sequential trial. This approach treats these decisions as part of the observed data, thus capturing all available information more accurately than traditional posteriors based solely on accumulated responses \citep{roberts1967informative}. Here, informative stopping rules refer to decisions taken at pre-planned interim analyses on the basis of accumulating data related to the primary outcome \citep[p.~88]{berger1988likelihood}.

Building on this, we extend the concept of conditioning on the realized design to multi-stage settings and introduce a novel metric to quantify how interim analyses in a GSD modify posterior beliefs about the treatment effect compared to a FSD.
As we will show, the same data can yield very different posterior estimates depending on whether informative adaptation is taken into account in the analysis, making it crucial to have a global measure of this discrepancy. This discrepancy is relevant because, in many applied analyses, posteriors are obtained without incorporating informative adaptive rules, thus overlooking that such rules can themselves convey information about the parameter \citep{flournoy2023posterior}.

The proposed metric is derived from the information-theoretic concept of entropy, which has seen wide application in Bayesian statistics, for example, to improve robust inference \citep{hooker2014bayesian,miller2019robust} or optimize resource allocation in experimental design \citep{mozgunov2019information,mozgunov2020information,caruso2024response,rainforth2024modern}. Since the realized sequence of decisions (e.g., whether to stop or continue at each interim look) depends on the observed data, the natural use of this metric is for post-hoc evaluation, quantifying how ignoring information from informative adaptations affects posterior inference and informing the design of future trials.  
However, we also propose a modification of this measure that can be used in the pre-experimental phase to guide trialists in selecting boundary options. In this context, we suggest using this additional metric to evaluate the expected impact of a given boundary configuration on the posterior inference, alongside more traditional considerations such as expected sample size and type-I and type-II error rates. 

Although the proposed measure is grounded in a Bayesian framework and requires specifying a prior distribution to analyze group sequential trial results, it imposes no specific requirements on the design type. That is, stopping boundaries do not need to be derived from criteria based on Bayesian posterior probabilities for the measure to be computed. This flexibility can be appealing to trialists more familiar with well-established frequentist boundaries, such as Pocock-type or O'Brien-Fleming-type. In any case, our main interest lies in how interim decisions impact treatment effect estimation, based on the collected data and chosen boundaries, regardless of the criterion used to determine these boundaries. In addition, Bayesian stopping criteria can often be expressed in terms of observed statistics exceeding certain thresholds, yielding decision rules that resemble frequentist thresholds, although formally derived from posterior probabilities.

The objective of this work is three-fold: (i) to better understand how interim analyses influence the estimation of the treatment effect, altering posterior beliefs compared to a FSD; (ii) to introduce an information-theoretic metric that practitioners can use to quantify the impact of this modification; and (iii) to provide insights on how the trial can be designed to mitigate these effects, offering practical guidance to trialists on making more informed choices regarding decision boundaries.

The remainder of the paper is structured as follows. In Section \ref{sec:uncondVScond}, we briefly review the two posterior formulations discussed by \cite{flournoy2023posterior}: the conditional-on-decision posterior, which conditions on the actual interim decisions made during the trial, and the unconditional posterior, which ignores informative stopping decisions [not to be confused with the frequentist use of ``unconditional" in \cite{berry1987interim}].
We illustrate how these two types of posterior distributions can differ substantially in a multi-stage GSD example with the possibility of early stopping for futility or efficacy. In Section \ref{sec:CoA_general}, we adopt an information-theoretic perspective to introduce a measure that quantifies the divergence of the conditional posterior distribution from the unconditional posterior distribution due to adaptation. We discuss the use of this metric for the post-hoc evaluation of a realized group sequential trial and show how this global measure relates to various discrepancies between conditional and unconditional posteriors, such as differences in estimated treatment effects and credible intervals. An expected version of this metric is presented in Section \ref{sec:ECoA} with the aim of using it in a pre-experimental context as a guidance for trialists in the choice of boundaries. An application of the proposed methodology is illustrated through a 3-stage clinical trial evaluating a treatment for central nervous system disorders, with the option of early stopping for efficacy. The paper ends with a discussion and suggestions for future developments in Section \ref{sec:discussion}.

\section{The consequences of ignoring informative adaptive decisions}
\label{sec:uncondVScond}

Consider a group sequential experiment with $S\geq2$ stages. Conditional on a decision to continue after stage $s-1$, a sample of $n_s$ independent and identical distributed observations ($\bm X_{s}$) is collected, with the single observation $X_i$ having density $f_{\theta}(x_i)$, where $\theta$ may be a scalar or vector-valued parameter. Suppose the objective is to evaluate a general estimand $\theta$ (e.g., mean, quantile or other function of the underlying distribution).  
Decisions at each interim $s$ are based on a statistic $T(\bm X_{(s)}, N_{(s)})$, where $N_{(s)}$ is the total sample size collected up to that point. For example, if $\theta = (\mu, \sigma^2)$ represents the mean and variance of a normal outcome, a natural choice for the statistic is the cumulative sample mean and variance at interim $s$, namely $T(\mathbf{X}_{(s)}, N_{(s)}) = (\bar X_{(s)}, S^2_{(s)})$, where $\bar X_{(s)} = N_{(s)}^{-1}\sum_{i=1}^{N_{(s)}} X_i$ and $S^2_{(s)} = (N_{(s)}-1)^{-1} \sum_{i=1}^{N_{(s)}} (X_i - \bar X_{(s)})^2$. More generally, $T$ can be any function of the accumulated data that summarizes the evidence about $\theta$. 

Let $L_{U_s}(\theta;\bm{x}_{(s)},)=\prod_{i=1}^{n_{(s)}}\,f_{\theta}(x_i)$ be the likelihood function for $\theta$ associated with the sample $\bm x_{(s)}=(x_1,\dots,x_{n_{(s)}})$ collected up to the end of stage $s$. This likelihood does not depend on any interim decisions made up to stage $s$. Following \cite{marschner2021general}, we refer to this as the unconditional likelihood to distinguish it from likelihoods that condition on the stopping decisions actually made, which will be discussed later. 

\subsection{Likelihood conditional on the interim decision made}
\label{sec:designLikelihood}

Let $D_s$ be a ternary variable denoting the informative decision to stop for efficacy ($D_s=1$), futility ($D_s=-1$) or to continue ($D_s=0$) the trial at interim analysis $s=1,\dots,S-1$. Let $\bm D_s=(D_1,\dots,D_s)$ be the vector of all informative interim decisions up to interim $s$. Due to the sequential design, the first $s-1$ values of the vector will be zeros and the last will depend on the decision made at interim $s$. Thus, the probability of observing a realization $\bm d_s$ of the design up to interim $s$ yields the following likelihood:
\begin{equation}
\label{eq:designLikelihood}
    \begin{aligned}
    L_{\bm D_s}(\theta; \bm d_s)&=P(\bm D_s=\bm d_s|\theta)=P\Biggl(\bigcap_{t=1}^{s-1}\{D_t=0\}\,\cap\,\{D_s=d_s\}\Biggr|\theta\Biggr)\\
    &=P(D_1=0|\theta)P(D_2=0|D_1=0,\theta)\cdots P(D_s=d_s|D_1=0,\dots,D_{s-1}=0,\theta)\\
    &=\prod_{t=1}^{s-1}\,P(D_t=0|D_{t-1}=0,\theta)\,\cdot\,P(D_s=d_s|D_{s-1}=0,\theta)\,,
    \end{aligned}
\end{equation}
with the convention that $P(D_0=0)=1$ and using the fact that $D_1,\dots,D_s$ is a Markov sequence, since $D_t$ is independent of $D_1,\dots,D_{t-2}$, given the value of $D_{t-1}$, for $t>2$. 

The expression in \eqref{eq:designLikelihood} is the likelihood of a design that has been realized up to interim decision $s$ \citep{marschner2021general}, and it properly accounts for the selection effect induced by all interim decisions that led to the stage $s$. In other words, the design likelihood reflects the fact that only the data patterns that resulted in the observed decision sequence could have been observed. Importantly, this likelihood does not involve the final decision made at the end of stage $S$, since this final stopping decision is fixed a priori.

Following \cite{marschner2021general}, the likelihood conditional on the realized design until interim $s$ is defined by the relationship $L_{U_s}(\theta;\bm x_{(s)})=L_{D_s}(\theta; \bm d_s) L_{C_s}(\theta;\bm x_{(s)}|\bm d_s)$, which can be interpreted as the total information arising from two independent sources: the interim decisions and the information in the sample conditional on the interim decisions.

\subsection{Bayesian posteriors}
\label{sec:posteriors}

Suppose that $\theta$ is provided with a prior distribution $\pi(\theta)$ and that a particular trajectory of informative interim decisions $\bm D_s=\bm d_s$ has been observed up to the end of the stage $s$. Then 
\begin{equation*}
    \pi_{U_s}(\theta|\bm x_{(s)})=\frac{\pi(\theta)L_{U_s}(\theta;\bm x_{(s)})}{\int\,\pi(\theta)L_{U_s}(\theta;\bm x_{(s)})\,d\theta}\propto\pi(\theta)L_{U_s}(\theta;\bm x_{(s)})\,,
\end{equation*}
is the unconditional posterior distribution for $\theta$ at the end of stage $s=1,\dots,S$. This posterior completely ignores interim looks that have occurred up to interim $s$ and only updates prior beliefs with data collected up to the end of stage $s$. Equivalently, this can be viewed as a ``counterfactual" posterior, namely the posterior we would have obtained had the data been collected under a a design without informative stopping decisions (e.g., FSD). In contrast, the posterior distribution, conditional on all the informative decisions made up to the end of stage $s$, is given by
\begin{equation}
\label{eq:condPost}
    \pi_{C_s}(\theta|\bm{x}_{(s)},\bm d_s)=\frac{\pi(\theta)L_{C_s}(\theta;\bm x_{(s)}|\bm d_s)}{\int\,\pi(\theta)L_{C_s}(\theta;\bm x_{(s)}|\bm d_s)\,d\theta}=\frac{\pi_{U_s}( \theta|\bm x_{(s)})}{B_{s}\cdot L_{D_s}(\theta; \bm d_s)}\,,
\end{equation}
where 
\begin{equation}
\label{eq:BF}
    B_{s}=\frac{\int\,\pi(\theta)L_{C_s}(\theta;\bm x_{(s)}|\bm d_s)\,d\theta}{\int\,\pi(\theta)L_{U_s}(\theta;\bm x_{(s)})\,d\theta}=
\int\,\frac{\pi_{U_s}(\theta;\bm x_{(s)})}{L_{D_s}(\theta; \bm d_s)}\,d\theta=\mathbb{E}_{\pi_{U_s}}[L_{D_s}^{-1}(\theta; \bm d_s)]\geq1\,,
\end{equation}
since $L_{D_s}(\theta; \bm d_s)\leq1$. Notice that $B_s$ is the Bayes factor \citep{kass1995bayes} between the conditional and the unconditional model. This quantity is always greater than $1$, showing better fit of the conditional model to the data, and it is equal to $1$ if and only if $L_{D_s}(\theta; \bm d_s)=1,\,\forall \theta$, which is the case where interim looks are completely irrelevant to the decision. 

While the unconditional posterior distribution treats the decision process as irrelevant to inference, reflecting the posterior beliefs based only on the observed data, the conditional posterior distribution reflects the posterior beliefs after explicitly accounting for the fact that the decision process is relevant to the final inference.

\subsection{Illustration: group sequential design with normally distributed data}
\label{sec:plotPosteriors}

\subsubsection{Posterior distributions with normal data}

Suppose that we conduct a group-sequential trial of some experimental treatment with up to $S$ analyses of a sample. Let the response for a generic patient $i$ be $X_i \sim N(\theta, \sigma^2)$, with known variance $\sigma^2$ and assuming that interim decisions are non-informative. If the prior distribution for $\theta$ is $N(\mu_0,\tau_0^2)$, the resulting unconditional posterior of $\theta$ at the end of stage $s=1,\dots,S$ is given by
\begin{equation*}
    \pi_{U_s}(\theta|\bar x_{(s)})\overset{d}{=}N\Biggl(\frac{\tau_0^2 \bar x_{(s)}+\frac{\sigma^2}{n_{(s)}}\mu_0}{\tau_0^2+\frac{\sigma^2}{n_{(s)}}},\,\frac{\tau_0^2\sigma^2}{n_{(s)}\tau_0^2+\sigma^2}\Biggr)\,.
\end{equation*}
Let $(f_1,\dots,f_s)$ and $(e_1,\dots,e_s)$ be the vectors of futility and efficacy stopping boundaries for the first $s$ interim analyses and let $g_{(s)}$ be the density of $s$ cumulative means, namely a multivariate normal distribution centered in $\bm \theta=(\theta,\dots,\theta)'$ with elements of the covariance matrix given by $Cov(\bar X_{(s_1)},\bar X_{(s_2)})=\sigma^2/n_{(s_2)}$ for each $s_1\leq s_2$. 
Then, the posterior distribution conditional on all the informative decisions made up to the end of stage $s$ is given by \eqref{eq:condPost},
with \eqref{eq:designLikelihood} being $$L_{D_s}(\theta; \bm d_s)=P(l_1\leq\bar X_{(1)}\leq u_1,\dots,l_s\leq\bar X_{(s)}\leq u_s)=\int_{l_1}^{u_1}\dots\int_{l_s}^{u_s}\,g_{(s)}(\bm t)\,d\bm t\,,$$ where $(l_t,u_t)=(f_t,e_t)$, for $t=1,\dots,s-1$, while the values of $l_s$ and $u_s$ depend on the decision made at interim $s$: if the trial is continued, then $(l_s,u_s)=(e_s,f_s)$; if it is stopped for efficacy, then $(l_s,u_s)=(e_s,\infty)$; if it is stopped for futility, then $(l_s,u_s)=(-\infty,f_s)$.

\subsubsection{Example: 3-stage group-sequential trial}
\label{sec:plotPosteriors_example}

For illustration, suppose that a group-sequential single-arm clinical trial of some experimental treatment with two interim analyses and a final analysis ($S=3$) is planned. We assume that the observed response $X_i$ for patient $i$ is normally distributed with mean $\theta$ and variance $\sigma^2=1$. At each interim analysis, efficacy and futility criteria are evaluated to decide if the trial should be stopped. At the final analysis, both efficacy and futility criteria are assessed, with the possibility of indeterminate outcomes when none of the two previous criteria is fulfilled \citep{bross1952sequential,gsponer2014practical}. We adopt \cite{o1979multiple}'s boundaries and $12$ patients per stage (maximum sample size: $n_{\max}=36$), which guarantees a maximum type-I error probability $\alpha=0.05$ under $H_0:\theta\leq0$ and a power $1-\beta=0.9$ under $\theta=0.5$. We assume a $N(\mu_0,\tau_0^2)$ prior distribution on $\theta$, with $\mu_0=0$ and $\tau_0=10\sigma/\sqrt{n_{\max}}=1.67$. This is a weakly-informative prior, as its impact on the unconditional posterior is ten times smaller than the final-stage data, helping isolate the effect of interim looks from prior choice. 

\begin{figure}
    \centering
    \includegraphics[width=1\linewidth]{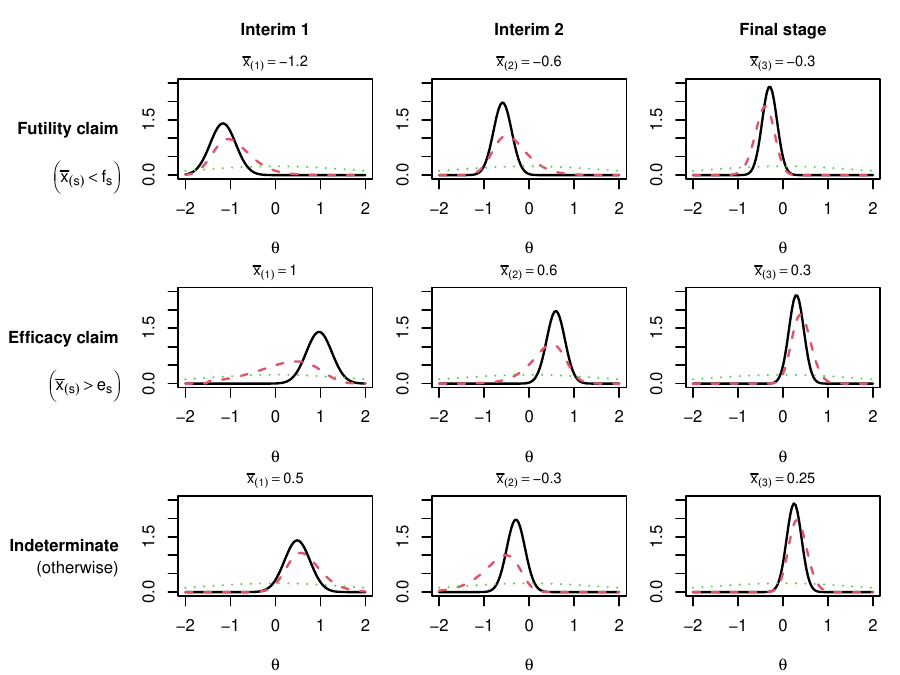}
    \caption{Unconditional (\textit{solid line}) and conditional (\textit{dashed line}) posterior distributions computed at interim and final analyses for different values of $\bar x_{(s)}$. Data collected at second interim and final analyses are conditional on having continued the trial up to that point. The vector of futility boundaries is $(-0.85, -0.43, -0.28)$, while efficacy boundaries are $(0.85, 0.43, 0.28)$. Prior distribution (\textit{dotted line}) used in the analysis phase is $N(0,1.67^2)$ and it is the same for all stages and scenarios.}
    \label{fig:3stagePost}
\end{figure}

Figure \ref{fig:3stagePost} shows conditional and unconditional posterior distributions for several scenarios, with different possible decisions occurring at the end of each stage. Unconditional posterior distributions at each stage have the same variance, that is affected only by the sample size, not by the observed summary statistics. In contrast, conditional posteriors vary much across scenarios within the same time. For example, observing a sample mean equal to either $0.5$ or $1$ at interim 1 (bottom and middle left panel, respectively) only shifts the mode of the unconditional posterior, while the conditional one also reflects the interim decision. This decision increases the posterior uncertainty on $\theta$ in both conditional posterior distributions, though to different extents. Specifically, stopping for efficacy at interim 1 after observing $\bar x_{(1)}>e_1$ yields a conditional posterior with much higher uncertainty, since $\bar x_{(1)}$ is close to the boundary $e_1$. 

Although the observed value is $\bar{x}_{(1)} = 1$, the resulting posterior distribution places most of its mass around values smaller than 1. This apparently counterintuitive outcome arises because, given that the trial stopped at interim 1, the observed $\bar{x}_{(1)}$ is more likely to have been generated by a smaller $\theta$. This effect is further reinforced by the fact that, had $\bar{x}_{(1)} < e_1$, a completely different decision would have been made. Thus, the conditional posterior down-weights larger values of $\theta$ accordingly.  

On the other hand, if we base our inference on the unconditional posterior - ignoring the fact that the trial stopped early - the resulting posterior estimates will tend to be overly confident about $\theta$ in the direction of the stopping decision. In other terms, the stopping rule acts as a filter: only certain data patterns allow the trial to stop early; and if we ignore this filtering effect using the unconditional posterior for our inference, we implicitly assume that weaker outcomes were still possible, even though they were already ruled out by the interim decision. As a result, the unconditional posterior in a sequential design with interim looks is artificially optimistic, as it fails to account for the fact that stopping depended on observing stronger early results. 

\begin{figure}
    \centering
    \includegraphics[width=0.8\linewidth]{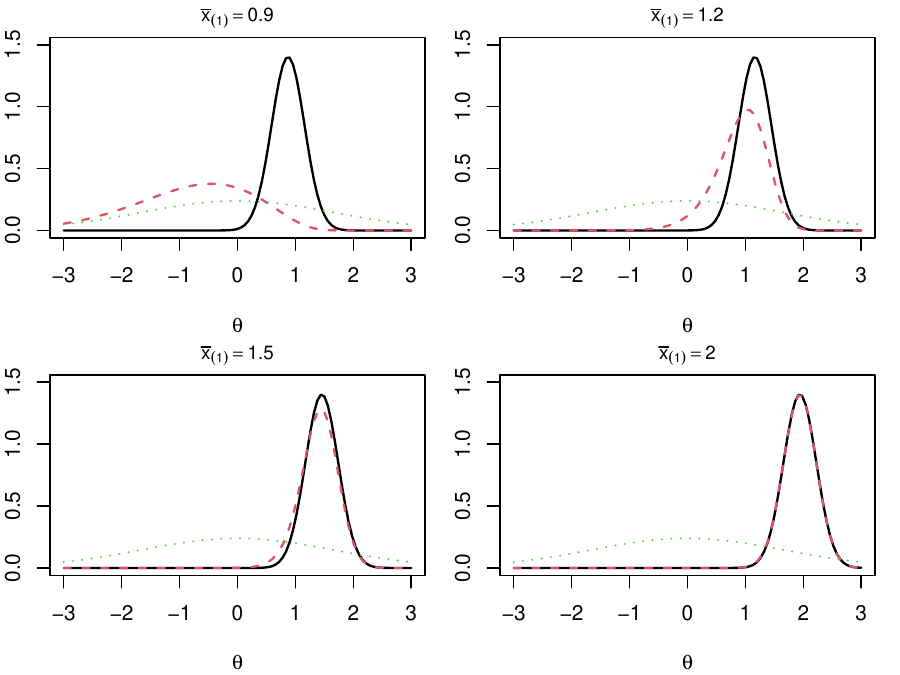}
    \caption{Unconditional (\textit{solid line}) and conditional (\textit{dashed line}) posterior distributions when trial is stopped for efficacy at the first interim for different values of $\bar x_{(1)}$ and efficacy boundary $e_1=0.85$. Prior distribution (\textit{dotted line}) used in the analysis phase is $N(0,1.67^2)$ and it is the same for all stages and scenarios.}
    \label{fig:int1Post}
\end{figure}

Figure \ref{fig:int1Post} offers a closer look at how the conditional posterior inference is affected when the trial stops for efficacy at interim 1. Intuitively, the closer $\bar{x}_{(1)}$ is to the threshold $e_1 = 0.85$, the flatter the conditional posterior of $\theta$ and the more its posterior mode shifts, reflecting the increased influence of the interim decision when estimating $\theta$. In particular, when $\bar{x}_{(1)} = 0.9$ - only slightly above the threshold $e_1 = 0.85$ - the decision to stop for efficacy is made with weaker evidence compared to when $\bar{x}_{(1)} = 2$, where the observed signal is much stronger.

\section{Measuring the impact of sequential adaptation on Bayesian posteriors}
\label{sec:CoA_general}

Figures \ref{fig:3stagePost} and \ref{fig:int1Post} provided insight on how posterior inference deviates from the FSD case when interim looks occur during the trial. In this section, we adopt the Kullback-Leibler (KL) divergence \citep{kullback1951information} to measure the effect of sequential monitoring in terms of posterior inference, capturing the degree of deviation of the conditional posterior distribution from the unconditional one. An introduction to KL divergence and the underlying information-theoretic concepts of surprisal and entropy is provided in the Appendix.

\subsection{KL divergence as measure of adaptation-induced posterior divergence}
\label{sec:CoA}

The KL divergence of the conditional posterior from the unconditional posterior,
\begin{equation}
\label{eq:CoA}
    D_{\bm d_s,\,\bm x_{(s)}}(\pi_{U_s}||\pi_{C_s}) = \int\,\pi_{U_s}( \theta|\bm x_{(s)})\log\frac{\pi_{U_s}( \theta|\bm x_{(s)})}{\pi_{C_s}(\theta|\bm{x}_{(s)},\bm d_s)}\,d\theta\,,
\end{equation}
quantifies the additional uncertainty on $\theta$ due to the use of an adaptive design with interim looks (which yields a conditional posterior) instead of a traditional FSD (which yields an unconditional posterior). We are interested in $D_{\bm d_s,\,\bm x_{(s)}}(\pi_{U_s}||\pi_{C_s})$ - rather than $D_{\bm d_s,\,\bm x_{(s)}}(\pi_{C_s}||\pi_{U_s})$ - as the unconditional posterior $\pi_{U_s}$ can be seen as the traditional distribution that we have when interim looks are not allowed or not relevant to the decision process. In other terms, it gives a measure of how the posterior beliefs on $\theta$ have been modified by the adaptation with respect to the FSD case. 
For this reason, we will refer to the metric in \eqref{eq:CoA} as adaptation-induced posterior divergence (AIPD). This metric can be alternatively written as 
\begin{equation}
\label{eq:CoA_JensenGap}
    D_{\bm d_s,\,\bm x_{(s)}}(\pi_{U_s}||\pi_{C_s})=
    \log\mathbb{E}_{\pi_{U_s}}[L_{D_s}^{-1}(\theta; \bm d_s)]-\mathbb{E}_{\pi_{U_s}}[\log L_{D_s}^{-1}(\theta; \bm d_s)]\,,
\end{equation}
namely as the difference between the logarithm of the average reciprocal of the design likelihood and the average of its logarithm, where the expectations are taken with respect to the unconditional posterior. This type of difference is often referred to as \textit{Jensen's gap} \citep{abramovich2016some}. The reciprocal of the design likelihood, $L_{D_s}^{-1}(\theta; \bm d_s)$, quantifies how surprising $\theta$ is given the realized design, in analogy with the definition of surprisal given in the Appendix. The first term of \eqref{eq:CoA_JensenGap}, $\log\mathbb{E}_{\pi_{U_s}}[L_{D_s}^{-1}(\theta; \bm d_s)]$, provides a global measure of how surprising the realized design is, averaged over all plausible values of $\theta$ according to the unconditional posterior distribution. The second term, $\mathbb{E}_{\pi_{U_s}}[\log L_{D_s}^{-1}(\theta; \bm d_s)]$, computes a local measure of surprise for each $\theta$, which is then averaged over the distribution of the unconditional posterior. Notably, the first term coincides with the Bayes factor between the conditional and unconditional models, which is a Bayesian measure of evidence in favor of the conditional model relative to the unconditional model. The higher the variability in how surprising the design is across different $\theta$, the stronger the evidence for conditioning on the realized design. This aligns with the intuition that while the adaptive design increases posterior uncertainty (e.g., wider credible intervals), it also makes the final inference more relevant to the particular design that was realized \citep{marschner2021general}.

Notably, since $L_{D_s}^{-1}(\theta; \bm d_s)>0$, the second-order Taylor expansion of $\log L_{D_s}^{-1}(\theta; \bm d_s)$ around $\mathbb{E}_{\pi_{U_s}}[L_{D_s}^{-1}(\theta; \bm d_s)]$ allows us to express the KL divergence in \eqref{eq:CoA_JensenGap} as approximately proportional to the ratio between the variance of the reciprocal of design likelihood and the square of its expected value, namely $D_{\bm d_s,\,\bm x_{(s)}}(\pi_{U_s}||\pi_{C_s})\approx \frac12\mathbb{V}_{\pi_{U_s}}(L_{D_s}^{-1}(\theta; \bm d_s))\,\mathbb{E}_{\pi_{U_s}}^{-2}[L_{D_s}^{-1}(\theta; \bm d_s)]\,.$
This relationship suggests that the KL divergence $D_{\bm d_s,\,\bm x_{(s)}}(\pi_{U_s}||\pi_{C_s})$ is directly related to how much the degree of surprise varies across different values of $\theta$ given the observed design realization. Notice that this measure of uncertainty is dimensionless as it reflects the relative dispersion of $L_{D_s}^{-1}(\theta; \bm d_s)$ around its mean. 

The quantity $D_{\bm d_s,\,\bm x_{(s)}}(\pi_{U_s}||\pi_{C_s})$ can be used to quantify the extent to which the interim looks modify the posterior distribution. The higher this quantity, the more affected is the final inference by the adapted design (e.g., higher uncertainty, larger credible intervals).

\subsection{Numerical illustration}
\label{sec:CoA_example}

\begin{figure}
    \centering
    \includegraphics[width=1\linewidth]{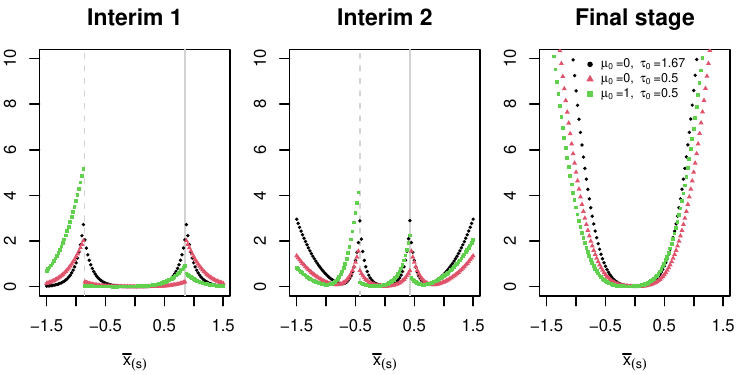}
    \caption{AIPD as function of the observed sample mean, $\bar x_{(s)}$, at interim and final analyses of a $3$-stage group sequential design. Vertical lines show informative efficacy (\textit{solid line}) and futility (\textit{dashed line}) boundaries at the two interim analyses. Different normal priors have been chosen: centered on a null treatment effect and either weakly-informative (\textit{points}) or moderately-informative (\textit{triangles}); moderately-informative and centered around a positive treatment effect (\textit{squares}).}
    \label{fig:KLwithXBarS}
\end{figure}

\begin{figure}
    \centering
    \includegraphics[width=1\linewidth]{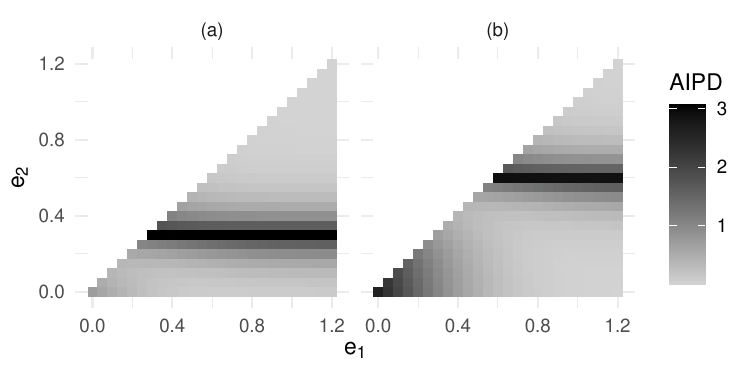}
    \caption{AIPD at interim 2 of a group sequential design with more than two stages for different combinations of the first two efficacy boundaries ($e_2\leq e_1$), given (a) $\bar x_{(2)}=0.3$ and (b) $\bar x_{(2)}=0.6$. Futility boundaries at interim 1 and 2 are $f_1=-0.85$ and $f_2=-0.43$. Each stage is composed by 12 patients and the prior distribution used in the analysis phase is $N(0,1.67^2)$.}
    \label{fig:KLwithE1E2}
\end{figure}

Following from the example of Section \ref{sec:plotPosteriors_example}, we illustrate the behavior of the metric introduced in \eqref{eq:CoA}. Figure \ref{fig:KLwithXBarS} shows the AIPD as function of the observed sample mean, $\bar x_{(s)}$, for different time points and prior specifications. At interim 1, the highest AIPD is attained for observed statistic values near the decision boundaries, reflecting the high uncertainty associated with the resulting decision. Similar findings were reported by \cite{tarima2024cost} in the context of a frequentist two-stage design, with the highest information loss occurring when the treatment effect was close to the stopping boundary. In contrast, strong evidence supporting the decision made (e.g., for $\bar x_{(1)}=-1.5$) leads to conditional posteriors that are nearly indistinguishable from their corresponding unconditional counterparts. Interestingly, at interim 1, high AIPD values are observed in cases of prior-decision conflict - that is, the prior provides overly optimistic support for the efficacy of the treatment, while data suggest its inefficacy. However, this conflict is resolved as the strength of support for the futility of the treatment increases, eventually overcoming the prior assumptions. AIPD values corresponding to observed statistic values that lead to trial continuation at interim 1 and 2 are not related to end-of-study scenarios. Nevertheless, they offer valuable insight into the impact of trial continuation on subsequent analysis time points. For example, large and small values of $\bar x_{(s)}$ at interim 2 and the final stage lead to high AIPD, reflecting the significant influence of earlier decisions to continue the trial on the conditional posterior distributions in subsequent analyses. 
Figure \ref{fig:KLwithE1E2} offers a closer look at the AIPD computed at interim 2 for different combinations of the first two efficacy boundaries, given the value of $\bar x_{(2)}$. The highest AIPD values occur when $e_2$ is close to the observed value of $\bar x_{(2)}$, or when both efficacy boundaries are much smaller than $\bar x_{(2)}$ (right panel), highlighting that continuing the trial beyond interim 1 may have introduced unnecessary distortion into the final analysis. In contrast, when both boundaries are much larger than the observed statistics $\bar x_{(2)}$, the decision to continue has little influence on the resulting conditional posterior distribution due to the high confidence in the interim decision. These findings underscore the importance of carefully selecting decision boundaries in future adaptive studies. For example, stopping a trial for efficacy based on limited evidence - such as when $\bar x_{(2)}$ only slightly exceeds $e_2$ - highlights the need of cautious  interpretation of the results and may motivate the adoption of more conservative efficacy boundaries in future studies to prevent premature conclusions.
Table \ref{tab:CoA} illustrates the relationship between AIPD and the discrepancies in posterior inference under conditional and unconditional posteriors in the same scenarios as those of Figure \ref{fig:3stagePost}. All the posteriors' summaries considered yield results consistent with AIPD across all considered scenarios. For example, the highest AIPD occurs when the trial is stopped for efficacy at interim 1, since $\bar x_{(1)}=1>e_1=0.85$. In this scenario, treating data from the GSD as if they came from an FSD  - and, consequently, making inference based on the unconditional posterior - leads to a significant overestimation of the treatment effect. This overestimation under the unconditional posterior arises from an excessive confidence in the decision to stop for efficacy, as it does not account for the uncertainty associated with this decision. For all scenarios considered, the overlap between the conditional and unconditional posterior credible intervals $95\%$ remains below $75\%$, and the proportion of the conditional posterior mass lying within the unconditional $95\%$ credible interval is always below $95\%$. This suggests that the intervals representing the most plausible treatment effects under the unconditional model are no longer as strongly supported under the conditional model. While each of these measures captures specific aspects of the two posterior distributions and their discrepancies, AIPD provides a global metric that integrates all these characteristics.
\begin{table}
\small
\caption{Adaptation-induced posterior divergence (AIPD), percentage of overlap between conditional and unconditional $95\%$ posterior credible intervals (OVCI\%), posterior conditional probability of $\theta$ being in the $95\%$ unconditional posterior credible interval (CPUI\%), ratio between conditional and unconditional posterior variances (PVR$_{C-U}$), and difference between conditional and unconditional posterior means ($\hat \theta_{ave}^C-\hat \theta_{ave}^U$) and modes ($\hat \theta_{max}^C-\hat \theta_{max}^U$) for several time-dependent scenarios of a $3$-stage GSD. Time of the analysis can be either interim 1 ($s=1$), interim 2 ($s=2$) or final analysis ($s=3$) and $\bar x_{(s)}$ refers to the time-specific sample mean. Decisions are based on the vector of futility boundaries $\bm{f}=(-0.85, -0.43, -0.28)$ and efficacy boundaries $\bm{e}=(0.85, 0.43, 0.28)$, with ``-" meaning indeterminate outcome. Rows highlighted in grey refer to scenarios that do not involve an end-of-study analysis.}
\setlength{\extrarowheight}{3pt}  
\setlength{\tabcolsep}{10pt}  
\begin{center}
\begin{tabular}{ccccccccc}
  \hline
 $s$ & $\bar x_{(s)}$ & Decision & AIPD & OVCI\% & CPUI\% & PVR$_{C-U}$ & $\hat \theta_{ave}^C-\hat \theta_{ave}^U$ & $\hat \theta_{max}^C-\hat \theta_{max}^U$ \\ 
  \hline
1 & -1.20 & Futility & 0.24 & 64.87 & 76.20 & 1.56 & 0.25 & 0.12 \\ 
1 & 1.00 & Efficacy & 1.04 & 31.98 & 39.37 & 2.60 & -0.86 & -0.52 \\ 
\rowcolor[gray]{.95}
1 & 0.50 & - & 0.16 & 70.53 & 81.67 & 1.42 & 0.18 & 0.08 \\ 
\hline
2 & -0.60 & Futility & 0.35 & 50.49 & 66.47 & 2.00 & 0.22 & 0.09 \\ 
2 & 0.60 & Efficacy & 0.35 & 50.91 & 66.47 & 1.99 & -0.23 & -0.09 \\ 
\rowcolor[gray]{.95}
2 & -0.30 & - & 0.66 & 43.25 & 53.76 & 2.24 & -0.42 & -0.22 \\ 
\hline
3 & -0.30 & Futility & 0.19 & 74.39 & 82.04 & 1.29 & -0.12 & -0.09 \\ 
3 & 0.30 & Efficacy & 0.19 & 72.77 & 82.04 & 1.30 & 0.12 & 0.09 \\ 
3 & 0.25 & - & 0.12 & 74.50 & 85.66 & 1.25 & 0.09 & 0.06 \\ 
   \hline
\end{tabular}
\end{center}
\label{tab:CoA}
\end{table}

\section{Expected adaptation-induced posterior divergence for pre-experimental decision-making}
\label{sec:ECoA}

In Section \ref{sec:CoA}, we introduced the KL divergence of conditional from unconditional posterior as a measure of the impact of adaptation in sequential trials on posterior inference. However, computing this measure requires observing the realized data under a specific design, making it more suitable for post-hoc evaluation of the chosen decision boundaries rather than as a pre-experimental tool for guiding their selection. 

If trialists seek a pre-experimental evaluation of the adaptation-induced posterior divergence at the end of the study, an expected version of this quantity can be derived. This involves integrating the AIPD over all possible trial realizations. Let $s^\star$ be the stopping stage, that is, either the interim phase at which a trial is stopped or the last planned stage of the trial. Although the AIPD can be computed at each stage of the trial, our primary interest is on the end-of-study AIPD, $D_{\bm d_{s^\star},\,\bm x_{(s^\star)}}(\pi_{U_{s^\star}}||\pi_{C_{s^\star}})$. For each stopping stage $s^\star$, we need to weight each possible $D_{\bm d_{s^\star},\,\bm x_{(s^\star)}}\bigl(\pi_{U_{s^\star}}\,\bigl|\bigl|\,\pi_{C_{s^\star}}\bigr)$ for the probability of observing that particular realization at stage $s^\star$. Thus, the expected end-of-study AIPD is given by
\begin{equation}
\label{eq:ECoA}
    \bar D(\theta) = \sum_{s^\star=1}^S\,P(S^\star=s^\star)\times\large\int\,D_{\bm d_{s^\star},\,\bm x_{(s^\star)}}\bigl(\pi_{U_{s^\star}}\,\bigl|\bigl|\,\pi_{C_{s^\star}}\bigr) \,d F_{\bm x_{(s^\star)}}(\bm x_{(s^\star)}|\theta)\,,
\end{equation}
where $P(S^\star=s^\star|\theta)$ is the probability of stage $S^\star=s^\star$ being the stopping stage and $F_{\bm x_{(s^\star)}}(\bm x_{(s^\star)}|\theta)$ is the probability distribution of the collected samples up to stage $s^\star$ (or of the $s^*$ statistics) whose stopping stage is $s^\star$. 
Although closed-form computation of \eqref{eq:ECoA} may be cumbersome, its value can still be approximated via Monte Carlo integration (more details are provided in the Appendix). 

Notice that $\bar D(\theta)$ offers a measure for pointwise evaluation of the expected AIPD, which can be adopted in cases where one is particularly interested in some particular scenario (e.g., $H_0:\theta=\theta_0$). If one wants a global measure over all the possible values of $\theta$, one can consider the area under the curve of $\bar D(\theta)$. 

\subsection{Numerical illustration: pre-experimental assessment of decision boundaries}
\label{sec:ECoA_example}

We consider a three-stage group-sequential trial to test a single dose of a drug to treat disorders of the central nervous system, which was described by \cite{dmitrienko2017analysis}. The primary objective of the trial is to demonstrate that the chosen dose of the drug was superior to a placebo in the acute treatment of subjects who meet the criteria for major depression. The efficacy of the experimental drug is evaluated using the mean reduction from baseline to the end of the 8-week study period in the total score of the 17-item Hamilton Depression Rating Scale (HAMD17). The standardized treatment effect is considered the primary endpoint, assuming that it is normal with mean $\delta$ and known standard deviation $\sigma=1$. 
Since the original analysis was not Bayesian, we consider a weakly-informative prior specification for $\delta\sim N(0,5)$, whose variance is $5$ times greater than one patient's response.

\begin{figure}
    \centering
    \includegraphics[width=1\linewidth]{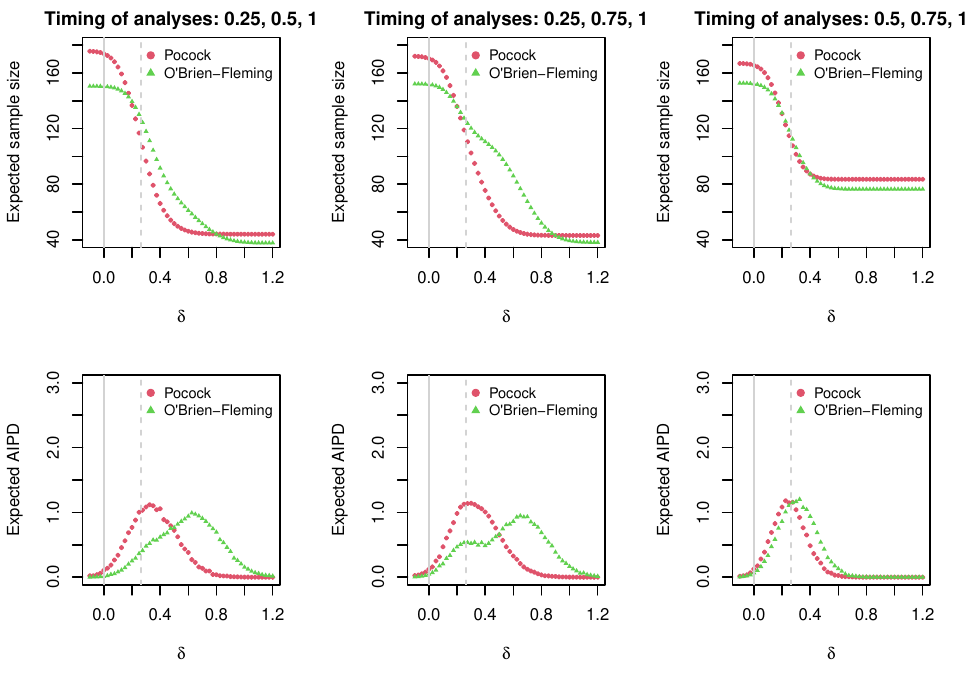}
    \caption{Expected sample size (\textit{top panels}) and expected end-of-study adaptation-induced posterior divergence (\textit{bottom panels}), $\bar D(\theta)$, when using Pocock-type and O'Brien-Fleming-type efficacy boundaries in a 3-stage trial with normal endpoint. Each column considers different schedules of interim analyses. 
    All boundary options guarantee a maximum type-I error probability $\alpha=0.025$ under $H_0:\delta\leq0$ (\textit{solid line}) and power $1-\beta=0.9$ under $H_1:\delta=0.265$ (\textit{dashed line}). Maximum sample size ranges from 167 to 176 for Pocock-type and 151 to 153 for O’Brien-Fleming-type boundaries across the three interim schedules.}
    \label{fig:ESSvsECoA}
\end{figure}

Figure \ref{fig:ESSvsECoA} shows the expected sample size and the expected end-of-study AIPD associated with Pocock-type and O'Brien-Fleming-type efficacy boundaries under three alternative schedules of interim analyses. For example, under the first schedule, the two interim analyses are conducted after roughly $25\%$ and $50\%$ of patients have been recruited, respectively. Pocock-type and O'Brien-Fleming-type efficacy boundaries are implemented via their corresponding $\alpha$-spending functions \citep{lan1983discrete} to accommodate for unequal spacing between interim analyses while maintaining $\alpha\leq0.025$ under $H_0:\delta\leq0$ and power $1-\beta=0.9$ under $H_1:\delta=0.265$. In all scenarios, O’Brien-Fleming-type boundaries are associated with both smaller expected sample size and smaller expected AIPD for small treatment effects ($\delta\leq0.2$), making them preferable when modest efficacy is anticipated. When the true treatment effect is large ($\delta>0.5$), Pocock-type boundaries lead to smaller expected sample size and expected AIPD, with the sweet spot being most evident when the first interim analysis is scheduled early (e.g., at $25\%$ of information). When the first interim analysis is delayed (e.g., $50\%$ of information), differences between the two boundary types are less sharp, with expected AIPD being negligible when $\delta>0.5$. Notably, the expected AIPD under O’Brien-Fleming-type boundaries is more sensitive to the interim schedule: for instance, earlier first looks tend to shift the peak of the expected AIPD towards larger values of $\delta$. The area under $\bar D(\theta)$, which averages performance across all possible $\delta$, is consistently smaller for Pocock-type boundaries (ranging between 13.8 and 19.5, compared with 15.9 to 22.2 for O'Brien-Fleming-type), suggesting that they reduce the expected divergence between conditional and unconditional posteriors. Overall, O’Brien-Fleming-type seems to be preferable when only modest treatment effects are expected, while Pocock-type boundaries are recommended when larger effects are more plausible or when global performance across $\delta$ is prioritised.

Notice that the expected AIPD represents an average across all the possible scenarios, given the value of $\delta$. If researchers are interested in exploring the variability of this measure, a boxplot can provide further insight about the distribution of the AIPD for each value of $\delta$.

\section{Discussion}
\label{sec:discussion}
 
In this paper, we have introduced a novel framework for quantifying the impact of informative interim decisions on Bayesian posterior inference in multi-stage group sequential designs. We have illustrated how informative stopping rules alter Bayesian estimates of the treatment effect and posterior credible intervals compared with fixed-sample designs, and have proposed a metric to quantify this discrepancy. This metric - adaptation-induced posterior divergence (AIPD) - arises naturally in a Bayesian setting but applies to both Bayesian or frequentist interim rules, thus bridging the two approaches. To our knowledge, no such measure has been previously proposed. We have also defined a pre-experimental version of the AIPD, enabling trialists to incorporate it at the design stage when selecting decision boundaries.

We have considered cases where stopping rules are informative; thus, ignoring whether interim data meet or do not meet stopping criteria would neglect the informativeness of the stopping process (\citealp{roberts1967informative}; \citealp[p.~88]{berger1988likelihood}). This highlights the key distinction between conditional-on-decisions and unconditional posteriors. The conditional posterior treats informative stopping decisions as part of the data. In contrast, the unconditional (or counterfactual) posterior considers only the observed responses, treating interim decisions as non-informative. This should not be confused with the frequentist “unconditional” inference of \cite{berry1987interim}, which averages over all possible experiments: consistent with the Bayesian framework of \cite{flournoy2023posterior}, we always condition on the events that actually occurred.

While we have focused on pre-planned informative adaptations, other unscheduled interim decisions can be practically informative. For example, in ophthalmic trials, stopping a study due to adverse events - such as vision deterioration, which may be closely related to the primary outcome - can provide relevant information for the parameter of interest \citep{friedman2025randomized}.

We have introduced the adaptation-induced posterior divergence (AIPD) as a global metric for quantifying the impact of these modifications in the posterior beliefs. This metric is grounded in the intuitive information-theoretic concepts of surprisal and entropy, acting as a measure of the additional uncertainty introduced by sequential monitoring that would not arise in a FSD. Ideally, an end-of-trial AIPD equal to $0$ means that the event of early stopping is almost impossible at each interim analysis, leading to a group sequential design which is nearly indistinguishable from a traditional fixed-sample design. This scenario facilitates a more straightforward comparison of treatment effect estimates between adaptive and non-adaptive designs. Smaller values of AIPD are preferable, as they indicate reduced distortion in the final analysis. However, apart from this, there is no natural benchmark for determining what can be considered a ``large" or ``small" divergence. This is a common issue with metrics based on KL divergence, that do not offer a direct statistical interpretation of the magnitude of the divergence. We suggest using the AIPD for comparing different scenarios, such as different observed samples or decision boundaries. For example, if one design yields a larger AIPD than another for the same data, this indicates that the interim decisions in that design have a greater influence on the posterior distribution. 
This measure can provide researchers with insight into the extent to which the interim analyses influence the final inference, and it can be useful in a post-hoc evaluation framework for several reasons. First, it allows them to assess how much the posterior distribution of the treatment effect deviates from what would have been obtained in a fixed-sample design. Second, they can compare different designs retrospectively by evaluating the trade-off between efficiency - enabled by the possibility of early stopping - and information loss across various adaptation strategies. For example, one adaptation strategy might allow for early stopping when statistical evidence emerges, leading to a more efficient trial. However, this efficiency gain could come at the cost of an increased bias in the treatment effect estimation, with the extent of bias depending on the uncertainty associated with the decision to stop. A different strategy might be more conservative and require stricter criteria for early stopping, ensuring more reliable posterior inference but requiring a longer trial duration. Third, it informs the design of future trials, allowing researchers to refine study protocols to mitigate undesirable distortions in the final inference.

The expected version of the AIPD, $\bar D(\theta)$, can be computed before a trial starts to assess whether a given design configuration is likely to lead to large distortions in inference. This measure varies with $\theta$ and, thus, it can be reported in a similar way as the expected sample size. Additionally, the area below the curve of $\bar D(\theta)$ gives a global summary of the expected AIPD across all the possible scenarios. This measure can help trialists in selecting alternative combinations of decision boundaries and interim schedules, based on the anticipated impact of interim decisions on posterior estimates.

For example, a natural extension of the current framework is to define a loss function that balances the trade-off between efficiency (how effectively the trial uses its sample size, as captured by ESS) and informativeness (the change in posterior precision due to informative decisions, quantified by the expected AIPD relative to a fixed-sample design). This loss could be evaluated across the parameter space and optimized subject to type-I error rate and power constraints, allowing the selection of stopping boundaries or other adaptive design parameters. Although such an implementation is beyond the scope of the present work, it opens a promising decision-theoretic avenue for future research.

To illustrate the concept behind our proposal, we have focused on the case of normal responses, an assumption commonly used in this type of analysis \citep{jennison1997group,stallard2020comparison}. For moderate to large sample sizes, the central limit theorem justifies the use of normal approximations for the stage-specific sample mean. However, for smaller sample sizes, we recommend considering more appropriate families of distributions.

In this paper, our interest has not focused on how decision boundaries are derived, but rather in understanding the impact of interim decisions on treatment effect estimation, given a chosen set of boundaries and the observed data. Although the proposed measure is rooted in a Bayesian framework and requires the specification of a prior distribution for analyzing the results of a group sequential trial, the design itself does not necessarily have to be Bayesian. That is, stopping boundaries are not required to be derived from criteria based on Bayesian posterior probability for the measure to be computed. Notably, frequentist GSDs can sometimes be seen as limiting cases of Bayesian designs based on posterior probability rules \citep{spiegelhalter1994bayesian}. \cite{stallard2020comparison} show that, when the same posterior probability threshold is applied at each interim analysis and equally spaced looks occur, a Bayesian design with a non-informative prior yields Pocock-type boundaries, while an unlikely, informative negative prior can mimic O’Brien-Fleming-type boundaries.
From a Bayesian point of view, distinct priors for the design and analysis phases are also possible \citep{de2007using}. For example, while the analysis-phase prior should reflect clinicians' prior beliefs about the treatment effect before the experiment, the design-phase prior can be chosen to achieve desirable boundary properties that resemble frequentist options.

Finally, we have focused on group sequential designs as the primary example of designs involving interim decisions because of their use in clinical trials. 
Future work might explore how these ideas can be extended to a wider range of adaptive designs, such as response-adaptive designs [e.g., \cite{hu2006theory}] or enrichment trials [e.g., \cite{frieri2023design}], providing valuable insights into the impact of interim decisions on posterior inference in various clinical trial contexts.

\section*{Acknowledgements}

This work was begun when G. Caruso was a visiting postdoctoral fellow at the Department of Statistics of George Mason University (Fairfax, VA, USA). He thanks the department for the hospitality. This report is independent research supported by the National Institute for Health Research (NIHR Advanced Fellowship, Dr Pavel Mozgunov, NIHR300576) and by UK Medical Research Council (MC\_UU\_00002/19). For the purpose of open access, the author has applied a Creative Commons Attribution (CC BY) licence to any Author Accepted Manuscript version arising.

\bibliographystyle{apalike}
\bibliography{biblio}


\clearpage
\appendix
\section*{Appendix}

\subsection*{A.1. Surprisal, entropy and Kullback-Leibler divergence}
\label{app:entropies}

Let $Y$ be a random variable with probability distribution $p(y)$. The negative log-transform of $p(y)$, $h_Y(y)=-\log p(y)=\log p^{-1}(y)$,
was introduced by \cite{shannon1948mathematical} as measure of the ``surprise" associated with the observation of the outcome $y$ from $Y$. This quantity is known in the statistical literature under different names, such as \textit{self-information}, \textit{information content} or \textit{surprisal} \citep{bartlett1952statistical,baldi2010bits,cole2021surprise,bickel2023statistical}. In the following, we will refer to this function with the term \textit{surprisal}, as we argue that it better aligns with the intuitive idea of how unexpected it is to observe the outcome $y$ from the random variable $Y\sim p_Y$. For example, if $Y\sim N(\mu,\sigma^2)$, it is more unexpected (or surprising) to observe the outcome $y=\mu+c$ ($c\in\mathbb{R}$) than $y=\mu$. 

The Shannon entropy, or expected surprise, associated with the probability distribution $p(y)$ is given by $H(p)=\mathbb{E}_{p(y)}[-\log p(y)]$. It is a measure of average surprise associated with the random variable $Y$, which can be used to summarize the degree of uncertainty and variability of $Y$ \citep{mackay2003information}. Notice that if the sample mean of $n$ i.i.d. normal random variables is $\bar Y_n\sim N(\mu,\sigma^2/n)$, its Shannon entropy $H(p_Y(y))=0.5+0.5\log(2\pi\sigma^2/n)$ is directly related to the variance of the single variable, $\sigma^2$, and the sample size $n$.

Similarly, if $p(y)$ and $q(y)$ are two possible probability distributions for $Y$, the cross-entropy of $q(y)$ relative to $p(y)$ is $H(p,q)=\mathbb{E}_{p(y)}[-\log q(y)]$, and it is a measure of the expected surprise when the distribution $q(y)$ is used instead of $p(y)$ to describe $Y$. Notice that $H(p,q)\geq H(p)$, and they are equal when $p\equiv q$. This suggests that the cross-entropy combines two sources of uncertainty on $Y$: (i) intrinsic uncertainty on $p(y)$ [given by $H(p)$] and (ii) additional uncertainty due to the use of $q(y)$ instead of $p(y)$. This second quantity is also known as the Kullback-Leibler (KL) divergence of $q(y)$ from $p(y)$\,, $D(p||q)=H(p,q)-H(p)=\mathbb{E}_{p(y)}\biggl[\log\frac{p(y)}{q(y)}\biggr]$.
This measure is asymmetric - generally $D(p||q)\neq D(q||p)$ - but is always non-negative and invariant under parameter transformations. 

\subsection*{A.2. Monte Carlo procedure to estimate the expected AIPD, $\bar D(\theta)$}

The following procedure can be adopted to estimate $\bar D(\theta)$ for each $\theta$:
\begin{enumerate}
    \item Simulate a large value $M$ of possible realizations of the random vector $\bar X_1,\dots,\bar X_S|\theta$. 
    \item For each realization, find the stopping stage $s^\star$, such that $s^\star=S$, in case of no early stopping, or $s^\star=\text{min}_{s=1,\dots,S-1}\{\,\bar x_s\notin (f_s,e_s)\}$, otherwise. Then compute $D_{\bm d_{s^\star},\,\bm x_{(s^\star)}}(\pi_{U_{s^\star}}||\pi_{C_{s^\star}})$.
    \item Average all values of $D_{\bm d_{s^\star},\,\bm x_{(s^\star)}}(\pi_{U_{s^\star}}||\pi_{C_{s^\star}})$ to obtain a Monte Carlo estimate of \eqref{eq:ECoA}.
\end{enumerate}

\label{lastpage}

\end{document}